\documentclass[12pt]{article}

\begin{document}
\vspace*{-.6in} \thispagestyle{empty}
\begin{flushright}
CALT-68-2292\\
CITUSC/00-044\\
hep-th/0008009
\end{flushright}
\baselineskip = 20pt

\vspace{.5in} {\Large
\begin{center}
Does Superstring Theory Have a Conformally Invariant
Limit?\end{center}}

\begin{center}
John H. Schwarz\footnote{Work supported in
part by the U.S. Dept. of Energy under Grant No.
DE-FG03-92-ER40701.}
\\ \emph{California Institute of Technology\\
Pasadena, CA  91125 USA}
\end{center}
\vspace{1in}

\begin{center}
\textbf{Abstract}
\end{center}
\begin{quotation}
\noindent This talk describes a proposal, due to Hull, for a
conformally invariant limit of superstring theory in six
dimensions.\end{quotation} \vspace{1in}

\centerline{Talk presented at the {\it Ninth Marcel Grossmann Meeting} (MG9)} \vfil

\newpage

\pagenumbering{arabic} 

\section{Introduction}

It is generally believed that string theory and M theory have no
dimensionless parameters, except those that arise as the vacuum
values of scalar fields. On the other hand, each of them seems to
have a fundamental length scale $\ell$, namely the string scale or
the eleven-dimensional Planck scale. In this brief talk, I would
like to draw your attention to a proposal, due to Hull
\cite{Hull}, to take $\ell \to \infty$. If such a limit exists, it
would send all masses to zero, and it should lead to a conformally
invariant theory. The resulting theory could be regarded as an
unbroken symmetry limit of string theory or M theory. As such, it
might be, in some sense, more fundamental. The specific proposal
of Hull, which I will describe in this talk, starts with the
maximally supersymmetric theory in five dimensions. Then,
according to the proposal, the limit $\ell \to \infty$ corresponds
to decompactification of a sixth dimension. In the limit one
should then be left with a conformally invariant theory in six
dimensions. This (conjectured) theory has a number of unusual
features. Clearly, more study is required to decide whether this
theory really exists, and if so to understand its properties
better.

\section{Dynamically Generated Dimensions}

As we have learned in recent years,
there are numerous instances where a dual viewpoint reveals additional compact dimensions.
The best-known example is Type IIA superstring theory in 10 dimensions. As Townsend and
Witten showed, when one goes beyond perturbation theory this is better viewed as M theory
compactified on a circle.  The radius of the circle is $R = \ell_s g_s$, where $g_s$ is the
string coupling constant and $\ell_s$ is the fundamental string scale. There are also more
complicated examples that exhibit similar features. For example, the heterotic string compacted
on a three-torus is dual to M theory compactified on K3. In this case the volume of the K3
is proportional to $\ell_s^4\, g_s$. A third example is a duality between heterotic string theory
compactified on a five-torus and the IIB superstring theory compactified on K3 times a circle.
In this case, the radius of the circle is given by $R = \ell_s g_s$, where $g_s$ is the heterotic
string coupling constant.  In each of these three examples, the size of compact dimensions is
dual to the string coupling constant, so that decompactification occurs in the strong-coupling
limit. Of course, the string coupling constant is determined as the vacuum value of
a scalar field (the dilaton). In the dual theory this field is identified as one of the
components of the metric, and there is no additional scalar field.

There is another example of a dynamically generated dimension that is a better
prototype for Hull's proposal. Consider maximally supersymmetric Yang--Mills theory
in five dimensions. This is non-renormalizable, of course, so one expects to
find new dynamics in the ultraviolet. In this case, it is known exactly what that
dynamics must be. Namely, there is a circular sixth dimension, whose radius is
determined by the Yang--Mills coupling constant $R= g_{YM}^2$. In the strong-coupling
limit the circle decompactifies and one is left with a superconformal field theory
in six dimensions. There are several considerations that support this interpretation.
For one thing, the five-dimensional theory has ``solitonic instantons" whose
mass is proportional to $g_{YM}^2$. Identifying these as Kaluza--Klein
excitations leads to the identification of KK charge with instanton number, and
gives the relation $R= g_{YM}^2$. A second way of thinking about this is
in terms of $N$ D4-branes in Type IIA superstring theory. Their world volume has
a maximally supersymmetric $U(N)$ Yang--Mills theory in five dimensions.
However, from the M theory viewpoint, a D4-brane is really an M5-brane with
one dimension wrapped on the circular dimension. Combining the D-brane
formula $g_{YM}^2 = g_s\ell_s$ with the relation $R= g_s \ell_s$, discussed in
the previous paragraph, again gives $R= g_{YM}^2$. In the limit one is left
with $N$ coincident M5-branes embedded in eleven-dimensional spacetime.
Their world volume theory is known from other considerations (such as AdS/CFT duality)
to be superconformal. The salient difference between this example and those of the
previous paragraph is that from the viewpoint of the 5d theory the coupling constant is
a dimensionful parameter of the theory, rather that the value of a scalar field.
In the limit, the only scale of the theory is removed,
which is crucial for the limiting theory to be conformal.
Eleven-dimensional M theory, by contrast,
has a scale and is not conformal.

\section{Supermultiplets in Five and Six Dimensions}

The little group that classifies massless particles in five dimensions is
$SO(3) \sim SU(2)$, whereas in six dimensions it is $SO(4) \sim SU(2) \times SU(2)$.
In terms of supermultiplets, when a D4-brane decompactifies to give an M5-brane,
a vector supermultiplet in five dimensions ($V_5$) turns into a tensor
supermultiplet in six dimensions ($T_6$). In terms of its $SU(2)$ content:
\begin{equation}
V_5 = [ {\bf 3} + 5\cdot {\bf 1} ] + [ 4\cdot {\bf 2} ],
\end{equation}
where we have grouped the eight bosons and eight fermions separately.
Similarly, the $T_6$ multiplet has the $SU(2) \times SU(2)$ content:
\begin{equation}
T_6 = [ {\bf (3,1)} + 5 \cdot {\bf (1,1)}] + [ 4 \cdot {\bf (2,1)}].
\end{equation}
The $T_6$ supermultiplet has $[2,0]$ supersymmetry, which means that both
supercharges have the same chirality. It should be contrasted with the $[1,1]$
vector supermultiplet
\begin{equation}
V_6 = [{\bf (2,2)} + 4 \cdot {\bf (1,1)} ] + [ 2 \cdot {\bf (2,1)} + 2 \cdot {\bf (1,2)} ],
\end{equation}
which can be obtained by dimensional reduction of a ten-dimensional vector supermultiplet.
It appears on the world volume of  type IIB 5-branes.

Maximal supergravity in six dimensions has $[2,2]$ supersymmetry. It can be regarded
as the low energy effective description of M theory compactified on a five-torus or of type IIB
compactified on a four-torus. The particle content can be obtained by forming
tensor products of the multiplets given above $SG_6 = V_6 \otimes V_6
= T_6 \otimes \tilde T_6$. By $\tilde T_6$ we mean the $[0,2]$ tensor supermultiplet with
reversed chirality. Either way, one finds that
\begin{equation}
SG_6 = [ {\bf (3,3)} + 5\cdot {\bf (3,1)} + 5\cdot {\bf (1,3)} + 16\cdot {\bf (2,2)}
+ 25\cdot {\bf (1,1)}]
\end{equation}
\[ +[4\cdot {\bf (4,1)} + 4\cdot {\bf (1,4)} + 24\cdot{\bf(2,1)} + 24 \cdot {\bf (1,2)}].
\]
As a check, note that there are 128 bosons and 128 fermions. The theory has U duality
group $SO(5,5;Z)$, and the 25 scalars take values in the moduli space
\begin{equation}
SO(5,5;Z) \backslash SO(5,5)/SO(5) \times SO(5).
\end{equation}
Reduction on another circle gives the five-dimensional supergravity multiplet
\begin{equation}
SG_5 = [ {\bf 5} + 27 \cdot {\bf 3} + 42 \cdot {\bf 1}] + [ 8 \cdot {\bf 4} + 48 \cdot {\bf 2}].
\end{equation}
This theory has U duality group $E_6(Z)$ and the 42 scalars take values in the moduli space
\begin{equation}
E_6(Z)\backslash E_{6,6} / USp(8).
\end{equation}

In addition to the $[2,2]$ supergravity multiplet $SG_6 = V_6 \otimes V_6
= T_6 \otimes \tilde T_6$, one can also form exotic supermultiplets with the
same amount of supersymmetry. Specifically, $V_6 \otimes T_6$ has $[3,1]$
supersymmetry and $T_6 \otimes T_6$ has $[4,0]$ supersymmetry. The particle
content in the latter case is
\begin{equation}
[ {\bf (5,1)} + 27 \cdot {\bf (3,1)} + 42 \cdot {\bf (1,1)}] +
[ 8 \cdot {\bf (4,1)} + 48 \cdot {\bf (2,1)}]. \label{fourzero}
\end{equation}
Note the similarity to the five-dimensional supergravity multiplet, now with all
the spin in the left $SU(2)$ factor. This is the same as the relationship between the supermultiplets $V_5$ and $T_6$.

\section{Hull's Proposal}

We are now in a position to state Hull's proposal. He suggests
that the five-dimensional theory (M theory compactified on a
six-torus or IIB theory compactified on a five-torus) has a
scaling limit $\ell \to \infty$ that results in a six-dimensional
theory with $[4,0]$ supersymmetry. In fact, $\ell$ should
correspond to the radius of a circular sixth dimension, much as
$g_{YM}^2$ did in the previous discussion. In the limit, all
excitations of the string theory, perturbative and
nonperturbative, become massless. So Hull conjectures that the
resulting theory is superconformal. Incidentally, a theory with
$[m,n]$ supersymmetry has an R-symmetry group $USp(2m) \times
USp(2n)$. Theories with $[3,1]$ or $[2,2]$ supersymmetry cannot be
superconformal, simply because there is no appropriate supergroup
containing these R symmetry groups.

One encouraging fact is that there is a plausible supergroup that
contains the $USp(8)$ R-symmetry group, namely $OSp(6,2|8)$, to
describe the superconformal symmetry. A striking feature of this
supergroup is that it contains 64 fermionic generators -- the 32
Poincar\'e supersymmetries plus 32 additional conformal
supersymmetries. We have not previously seen any sensible theories
with more than 32 fermionic generators (in my opinion), so
according to taste, this is either a remarkable possibility or a
cause for concern. The complete massless sector in six dimensions
is not easily described, but it should contain the $[4,0]$
supermultiplet of the previous section. Moreover, the $E_6(Z)$ is
present for all $\ell$, so it seems reasonable to conjecture that
it is also an exact symmetry in the limit. 

One of the striking features of the $[4,0]$ supermultiplet (\ref{fourzero})
is that it does not contain a conventional graviton. A graviton has spin ${\bf (3,3)}$,
whereas the $[4,0]$ supermultiplet contains ${\bf (5,1)}$ instead. On reduction to
five dimensions this gives the graviton, but no accompanying Kaluza--Klein vector
or scalar. This fact has several implications. For one thing, the theory
cannot be based on conventional Riemannian geometry. It is an interesting challenge to
figure out what replaces it.  Compactification on a circle gives an effective five-dimensional
theory that does have a conventional geometric description. What is unusual is that
the radius of the circle is not a modulus (there is no corresponding scalar field),
rather it defines a fundamental scale. Also, as there is no Kaluza--Klein gauge field
associated to the circle, the KK conserved charge stems from topology rather than
coupling to a gauge field.

Note that the limiting theory seems to have an infinite massless
spectrum. This may seem disturbing, but it is analogous to the
strong coupling limit of five-dimensional super Yang--Mills. The
superconformal theory associated with $N$ coincident M5-branes
also involves {\it tensionless strings}. As in that case, one does
not expect to be able to describe the theory by an explicit action
based on a local Lagrangian density. Hull refers to the mysterious
interactions of theories like this as ``M interactions''. So, he
argues, this theory also has M interactions. As we have learned
from the study of other conformal field theories, the physically
meaningful quantities to compute are correlation functions of
gauge invariant operators. It might be worthwhile to think about
this theory from that viewpoint.

The main reason we have learned so much about conformal field
theories in recent years is through AdS/CFT duality. When one
proposes a new CFT it is now natural to ask whether it might have
an AdS dual. Hull has considered the possibility that this might be the case for this
theory. However, this would seem to require a theory defined on an
AdS${}_7$ space times a compact space whose isometries account for
the $USp(8)$ R symmetry. Moreover, this theory would need to have
64 linearly realized supersymmetries. To me, this sounds like a
tall order. However, one can reject this, and still entertain the
possibility that the rest of the story is true.

\section{A Possible Problem}

Hull's theory, compactified on a circle, corresponds to M theory
on a six-torus, and has the U duality group $E_6(Z)$. Similarly,
Hull's theory compactified on a two-torus should correspond to M
theory compactified on a seven-torus, and should therefore have
the U duality group $E_7(Z)$. However, $E_7$ does not contain $E_6
\times SL_2$ as a subgroup. This suggests that Hull's theory does
not give the usual $SL(2,Z)$ modular group for the
two-torus.\footnote{This arose in a discussion with E. Witten.} It
seems that are two possibilities (both of which were suggested to
me by Hull). Either the $SL(2,Z)$ duality is not required, because
Hull's theory does not have conventional geometry, or the theory
in four dimensions has an additional $SL(2,Z)$ duality that has
not been noticed. Either way we would learn something very
interesting.

\section{Conclusion}

We have long suspected that string theory should have a limit in
which all gauge symmetries are unbroken. Hull's theory, if it
exists, would be a candidate to play this role. A natural question
in this regard is to identify what the unbroken gauge group is. At
the least, it should incorporate all the gauge symmetries
associated with all the excited states of all the different
strings in five dimensions, whose charges form a {\bf 27} of
$E_{6,6}$. Maybe there is a nice answer.

One important issue is the origin
of the Kaluza--Klein excitations from the five-dimensional
viewpoint. In other words, what is the analog of D0-branes in type
IIA superstring theory? Hull has identified the 5d charge that becomes
the KK momentum of the six-dimensional $[4,0]$ theory and shown
that it is carried by
gravitational-instanton solitons. Indeed, U duality implies that states
carrying these charges must be present in five dimensions. These states
still need to be better understood.

Hull's proposal seems to me to be potentially very important. The
question that needs to be answered first is whether the theory
exists. I have not yet formed a strong opinion one way or the other. 

\section*{Acknowledgments}

I am grateful to E. Witten and C. Hull for discussions.

\end{document}